\documentclass[epj]{webofc}
\usepackage[utf8]{inputenc}
\usepackage[varg]{txfonts}   
\usepackage{booktabs}
\usepackage{xcolor}
\definecolor{darkred}{rgb}{0.4,0.0,0.0}
\definecolor{darkgreen}{rgb}{0.0,0.4,0.0}
\definecolor{darkblue}{rgb}{0.0,0.0,0.4}
\usepackage[bookmarks,linktocpage,colorlinks,
    linkcolor = darkred,
    urlcolor  = darkblue,
    citecolor = darkgreen]{hyperref}
\usepackage{subfigure}
\wocname{EPJ Web of Conferences}
\woctitle{Lattice2017}
 \newcommand{\be}{\begin{equation}}
\newcommand{\ee}{\end{equation}}
 \newcommand{\bel}{\begin{align}}
\newcommand{\eel}{\end{align}}
\newcommand{\bea}{\begin{eqnarray}}
\newcommand{\eea}{\end{eqnarray}}

\begin{document}
%
\selectlanguage{english}
\title{%
$N\pi$ scattering in the Roper channel}
\author{%
\firstname{M.} \lastname{Padmanath}\inst{1}\fnsep\thanks{Speaker \email{Padmanath.M@physik.uni-regensburg.de} }, 
\firstname{C. B.} \lastname{Lang}\inst{2}, 
\firstname{Luka}  \lastname{Leskovec}\inst{3} \and
\firstname{Sasa}  \lastname{Prelovsek}\inst{1,4,5}
}
\institute{%
Instit\"ut  f\"ur  Theoretische Physik, Universit\"at Regensburg, D-93040 Regensburg, Germany
\and
Institute of Physics, University of Graz, A-8010 Graz, Austria
\and
Department of Physics, University of Arizona, Tucson, AZ 85721, USA
\and
Faculty of Mathematics and Physics, University of Ljubljana, 1000 Ljubljana, Slovenia
\and
Jozef Stefan Institute, 1000 Ljubljana, Slovenia
}
\abstract{%
We present results from our recent lattice QCD study of $N\pi$ scattering in the positive-parity
nucleon channel, where the puzzling Roper resonance $N^*(1440)$ resides in experiment. Using a variety
of hadron operators, that include $qqq$-like, $N\pi$ in $p$-wave and $N\sigma$ in $s$-wave, we systematically
extract the excited lattice spectrum in the nucleon channel up to 1.65 GeV. Our lattice results indicate that
N$\pi$ scattering in the elastic approximation alone does not describe a low-lying Roper.  Coupled channel
effects between $N\pi$ and $N\pi\pi$ seem to be crucial to render a low-lying Roper in experiment, reinforcing 
the notion that this state could be a dynamically generated resonance. After giving a brief motivation for studying 
the Roper channel and the relevant technical details to this study, we will discuss the results and the conclusions 
based on our lattice investigation and in comparison with other lattice calculations.}
\maketitle
\section{Introduction}\label{intro}
Understanding the baryon excitations is crucial to enhance our access to strong interactions 
in the low energy domain. One of the multiple interesting light baryon excitations, is the $N^*(1440)$ 
resonance  - the first excitation of nucleon, also known as Roper resonance with $(I)J^P=(1/2)1/2^+$.  
It has been elaborately studied and discussed in literature, but continues to be puzzling since 
its discovery in 1964~\cite{Roper:1964zza}. This resonance predominantly decays to $N\pi$, 
while decay modes such as $N\pi\pi$, $N\rho$ and $\Delta\pi$ are also observed collectively 
to a significant fraction. 

The most striking feature of this resonance is that it appears lower in mass with respect
to $N^*(1535)$, the lowest negative parity excitation of the nucleon with $(I)J^P=(1/2)1/2^-$. This 
hierarchy is in contrast with results from most phenomenological studies that assumed the
$N^*(1440)$ resonance to be of three quark nature \cite{Isgur:1978xj,Capstick:1986bm}. A variety 
of phenomenological suggestions that go beyond the  three quark picture
 followed these puzzling observations \cite{Crede:2013sze}.

There has been multiple lattice studies of the excited nucleon spectrum as well, most 
of which indicated an inverted hierarchy with respect to the experiments 
\cite{Liu:2014jua,Alexandrou:2013fsu,Alexandrou:2014mka,Engel:2013ig,Edwards:2011jj,Mahbub:2013ala,Roberts:2013ipa}. 
Unlike other lattice investigations, Ref. \cite{Liu:2014jua} observed a low lying Roper resonance using fermion 
discretization with good chiral properties and different eigenenergy extraction techniques. The question 
on the importance of the role of chiral symmetry remains to be confirmed from future calculations. All
these previous lattice calculations utilized three-quark interpolating fields (assuming $N^*(1440)$ to 
be stable) and identified the first excited lattice level with $N^*(1440)$. A recent lattice study that 
included five quark interpolating fields used strictly local $qqqq\bar q$ interpolators \cite{Kiratidis:2016hda}. 
However, either of these kinds of interpolating fields could suffer from lack of coupling with strongly decaying broad 
resonance, such as $N^*(1440)$. 

In this talk, we present the results from our recent lattice investigation of the excited nucleon spectrum within 
an energy region below 1.65 GeV, where the Roper resonance is observed \cite{Lang:2016hnn}. In order to determine 
the complete discrete spectrum we included for the first time $N\pi$ in $p$-wave as well as $N\sigma$ in $s$-wave 
in order to account for their scattering, along with $qqq$ type interpolating fields. We investigate the possible 
description of the Roper resonance based on our lattice results within elastic approximation of $N\pi$ scattering. 
In Section \ref{sec-2}, we briefly describe the methods and technicalities in our calculation. We discuss the results 
in Section \ref{sec:results} and then we summarize.

\section{Lattice methodology}\label{sec-2}
We investigate the excited nucleon spectrum on an $N_f=2+1$ ensemble generated by the PACS-CS collaboration 
with lattice extension $V=32^3\times 64$, physical volume $L^3\simeq (2.9~$fm$)^3$ and $m_\pi=156(7)(2)~$MeV
\cite{Aoki:2008sm}. The valence and the sea quarks are realized using non-perturbatively improved 
Wilson-clover fermions. 

\textbf{Quark smearing width and distillation} :  
For the  valence quarks sources and propagators we use full distillation \cite{Peardon:2009gh} with two different 
smearing widths (narrow [n] and wide [w] using 48 and 24 eigenvectors of the discretized gauge-covariant Laplacian). 
This technique aids us to build all the required Wick contractions, which are quite cumbersome for this study 
involving as many as 113 distinct diagrams. Using two smearing widths we hope to get enhanced access to form radial 
nodes in the wave function, which is favourable for the Roper resonance \cite{Roberts:2013ipa}. 

\textbf{Interpolators} :
We employ the following nucleon (${\cal N}^i$), pion ($\pi$) and sigma meson ($\sigma$) operators in the dispersion relation 
studies as well as in constructing the basis of baryon and baryon-meson interpolators used in extracting the excited 
nucleon spectrum. 
\begin{align}
\label{N}
{\cal N}^i_\mu(\mathbf{n})&\!=\!\sum_{\mathbf{x}} \epsilon_{abc} [u^{aT}(\mathbf{x},t)  \Gamma_2^i d^b (\mathbf{x},t)] ~[\Gamma_1^i q^c(\mathbf{x},t)]_{\mu}~\mathrm{e}^{i\mathbf{x\cdot n}\frac{2\pi}{L}}, \\
\label{pi}
\pi^+(\mathbf{n})& =\sum_{\mathbf x} \bar d({\mathbf x},t)\gamma_5 u({\mathbf x},t) \mathrm{e}^{i\mathbf{x\cdot n}\frac{2\pi}{L}},  \\
\pi^0(\mathbf{n}) &=\tfrac{1}{\sqrt{2}}\sum_{\mathbf x} [\bar d({\mathbf x},t)\gamma_5 d({\mathbf x},t)-\bar u({\mathbf x},t)\gamma_5 u({\mathbf x},t)] \mathrm{e}^{i\mathbf{x\cdot n}\frac{2\pi}{L}},\nonumber \\
\label{sigma}
\sigma(0)&=\tfrac{1}{\sqrt{2}}\sum_{\mathbf x} [\bar u({\mathbf x},t)u({\mathbf x},t)+\bar d({\mathbf x},t) d({\mathbf x},t)]~.
\end{align}   
The proton field is given by ${\cal N}^i_\mu(\mathbf{n})|_{q=u}$, whereas the neutron is given by ${\cal N}^i_\mu(\mathbf{n})|_{q=d}$. 
With three different choices of $(\Gamma_1^i,\Gamma_2^i)=(\mathbf{1},C\gamma_5),~(\gamma_5,C),~(i\mathbf{1},C\gamma_t\gamma_4)$, we employ 
three nucleon fields in this study. 

With these single hadron fields, we construct the following 10 baryon and baryon-meson annihilation operators with $(I)J^P=(1/2)1/2^+$ 
and total momentum zero. 
\begin{align}
O_{1,2}^{N\pi}&=-\sqrt{\tfrac{1}{3}} ~\bigl[p^{1,2}_{-\frac{1}{2}}(-e_x) \pi^0(e_x)-p^{1,2}_{-\frac{1}{2}}(e_x) \pi^0(-e_x)
-i p^{1,2}_{-\frac{1}{2}}(-e_y) \pi^0(e_y)+i p^{1,2}_{-\frac{1}{2}}(e_y) \pi^0(-e_y)\nonumber\\
& \qquad+ p^{1,2}_{\frac{1}{2}}(-e_z) \pi^0(e_z)-p^{1,2}_{\frac{1}{2}}(e_z) \pi^0(-e_z)\bigr]
  +\sqrt{\tfrac{2}{3}} ~\bigl[\{p\to n, \pi^0\to \pi^+\}\bigr] \quad [n], \nonumber\\
O_{3,4,5}^{N_w}&=p^{1,2,3}_{\frac{1}{2}}(0)\quad [w], \quad\quad  O_{6,7,8}^{N_n}=p^{1,2,3}_{\frac{1}{2}}(0)\quad [n], \quad\quad 
O_{9,10}^{N\sigma}=p^{1,2}_{\frac{1}{2}}(0) \sigma(0) \quad [n]. 
\label{O}
\end{align} 
Here $e_x$, $e_y$, and $e_z$ refer to the unit lattice momentum vectors along $x, y$, and $z$ directions, 
$n$ and $w$ inside square brackets refer to narrow and wide smearing respectively and the subscripts 
$\pm\frac12$ to the proton fields refer to the spin projections. The basis in eqn (\ref{O}) contains all the baryon-meson 
interpolators with non-interacting levels below $1.65~$GeV. 

The $\sigma$ meson is a broad resonance and can decay into 2$\pi$; $O^{N\sigma}$ is expected to couple to 
the $N\sigma$ as well as $N(0)\pi(0)\pi(0)$ levels. In this work, we implemented only the $O^{N\sigma}$ interpolator 
with the expectation that it will effectively represent a mixture of $N\sigma$ and $N(0)\pi(0)\pi(0)$.

\textbf{Extracting the spectrum} :  
Expanding the correlation matrix, 
\be
C_{ij}(t)=\langle \Omega|O_i(t+t_{src})\bar O_j (t_{src})|\Omega\rangle = \sum_n \langle \Omega | O_i|n\rangle \mathrm{e}^{-E_nt} \langle n|\bar O_j |\Omega\rangle = \sum_n  Z_i^n Z_j^{n*} \mathrm{e}^{-E_nt}
\label{C}
\ee
one can see that the large-time behavior of the eigenvalue  $\lambda^{(n)}(t,t_0)$ provides $E_n$, 
where as the eigenvectors $u_i^{(n)}$ are related to the operator overlaps $Z_i^n$. The spectrum is determined by performing non-linear fits to the eigenvalues $\lambda^{(n)}(t)$ 
extracted from the generalized eigenvalue problem \cite{Michael:1985ne,Luscher:1985dn} for the correlation matrices $C_{ij}(t)$
\be
\label{gevp}
 C_{ij}(t)u_j^{(n)}=\lambda^{(n)}(t,t_0)C_{ij}(t_0)u_j^{(n)}\;,\ \  \lambda^{(n)}(t,t_0)\propto  \mathrm{e}^{-E_n (t- t_0)}
\ee

\begin{figure}[bt] %
\centering
\sidecaption
\includegraphics[width=72mm,clip]{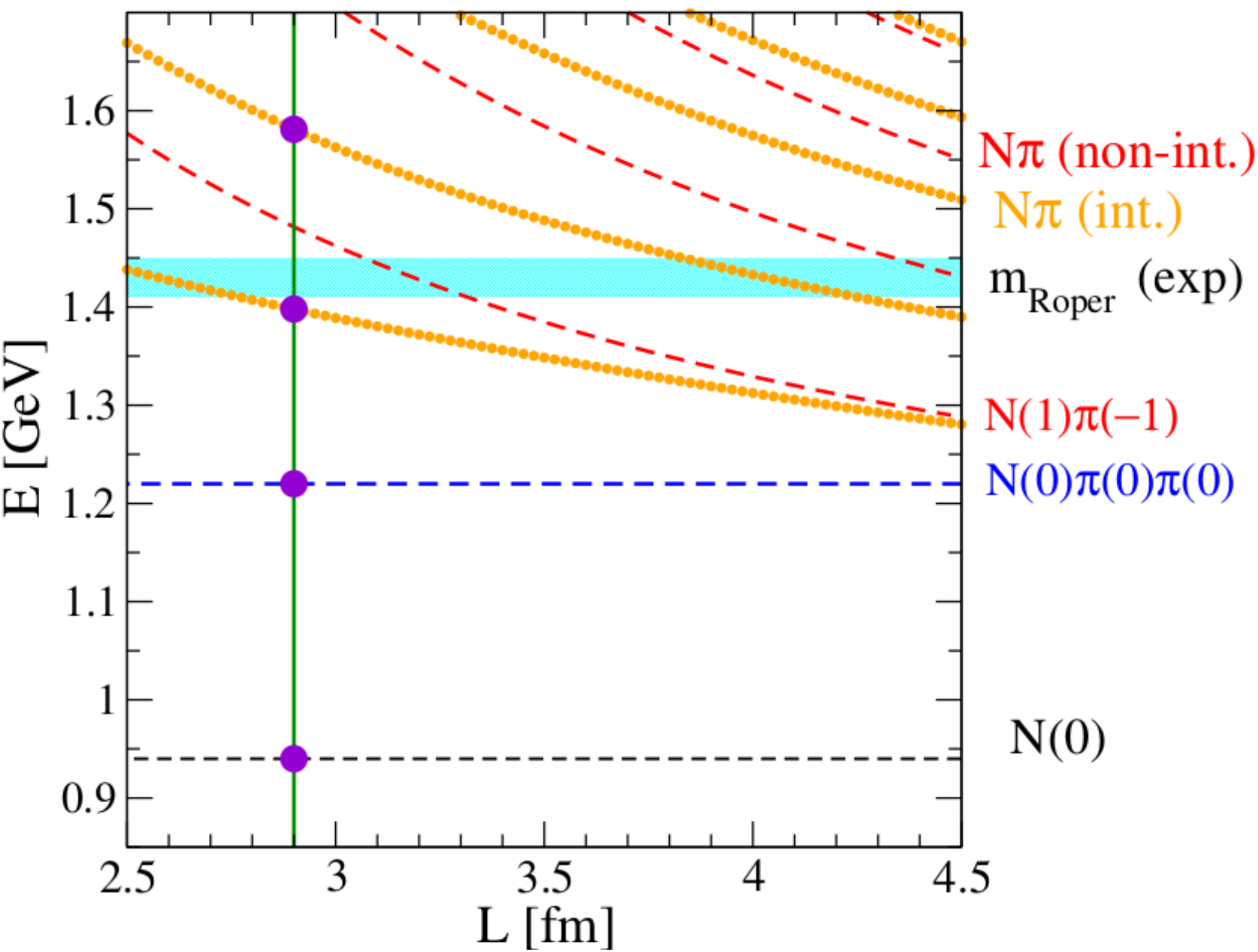}
\caption{Lattice spectrum as a function of the lattice extension L, predicted by eqn. (\ref{luscher}) within 
elastic $N\pi$ scattering. The experimental mass of the Roper (cyan band) and experimental phase shift information 
have been used as inputs to determine the energy shifts on the lattice. $N(0)\pi(0)\pi(0)$ (dashed blue line) 
and $N(0)\sigma(0)$ are assumed to be decoupled with the Roper and hence remain unaffected. }
\label{elasNpi}
\end{figure}

\textbf{Extracting the resonance information} : In a scenario, when there is no meson-meson and baryon-meson interactions, 
we expect to observe levels corresponding to $N(0)$, $N(0)\pi(0)\pi(0)$ and $N(1)\pi(-1)$ on the ensemble 
we use. With non-trivial attractive interaction the extracted levels get affected and thus the observed levels and their 
respective energies will have implications in the presence of resonances. Within the approximation of elastic $N\pi$ 
scattering, the interacting $N\pi$ levels (dotted yellow lines) gets shifted with respect to the respective non-interacting 
positions (dashed orange lines) as shown in Fig. \ref{elasNpi}. The figure predicts the lattice spectrum as a function 
of the lattice extension L given the experimental mass of the Roper (cyan band) and experimental phase shift information 
as inputs to L\"uscher's relation  \cite{Luscher:1990ux,Luscher:1991cf}  
\be
\label{luscher}
\delta(p)=\mathrm{atan}\biggl[\frac{\sqrt{\pi} p L}{2\,Z_{00}(1;(\tfrac{pL}{2\pi})^2)}\biggr],\  E_{n.i.}=E_{N(p)}+ E_{\pi(p)}
\ee   
where $E_{H(p)}$ is the energy of the hadron. 
The numerically extracted lattice spectrum from the correlation matrices we compute is compared with these analytic 
predictions and explored in search of a possible description of $N^*(1440)$ feature as a conventional resonance
within the elastic approximation of $N\pi$ scattering.

\section{Results and discussion} \label{sec:results}

\begin{figure}[htb] %
\centering
\sidecaption
\includegraphics*[height=63mm,clip]{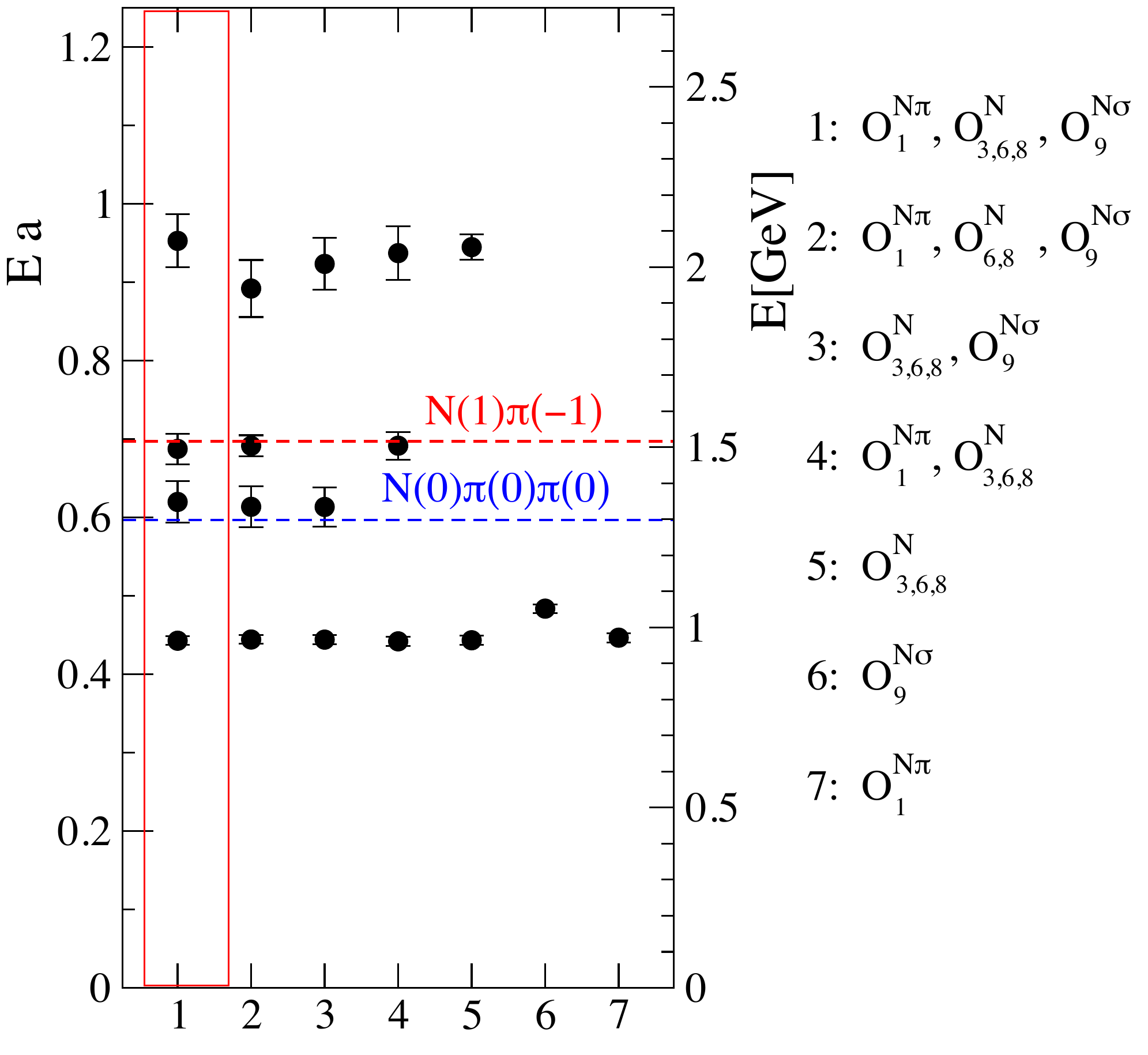} 
\caption{Lattice excitation spectrum of the nucleon for various choices of the interpolator basis. 
Basis 1 is referred to as the complete set, as it contains all types of interpolators and 
is highlighted with a red rectangle in the figure. The non-interacting scattering levels 
are shown as dashed horizontal lines (red for $N(1)\pi(-1)$ and blue for $N(0)\pi(0)\pi(0)$).}
\label{fig:E_interp_set}
\end{figure}

In Fig. \ref{fig:E_interp_set}, we present our results for the excited nucleon spectrum. These are 
based on correlation matrices for subsets of the operators basis 
$(O_1^{N\pi},~ O^{N_n}_3,~O^{N_w}_{6,8},~O_{9}^{N\sigma})$ as shown in the Fig. 
\ref{fig:E_interp_set} legend. Basis 1 (within the red rectangle) represents the set with all 
five operators in the above basis, which we call the complete set containing all types of 
interpolators.  Including the rest of the operators in  (\ref{O}) makes the spectrum 
noisier. The horizontal dashed blue and red lines denote the multi-hadron levels $m_N+2m_\pi$ and 
$E_{N(1)}+E_{\pi(-1)}$ in the non-interacting limit.
     
The main observation is that in basis 1 there are only three levels  below $1.65~$GeV. The ground 
state represents the nucleon and the remaining two levels lie close to the non-interacting 
positions of $m_N+2m_\pi$ and $E_{N(1)}+E_{\pi(-1)}$. As expected the ground state 
representing the nucleon can be seen to be quite stable with respect to different bases used, 
except for basis 6. In basis 6 only $O_{9}^{N\sigma}$ is used and hence the extracted level 
could be a mixture of $N$, $N(0)\pi(0)\pi(0)$ and $N(0)\sigma(0)$, since the $\sigma$ meson can 
also mix with the vacuum. The first excited state in basis 5, where only interpolators of type $qqq$ 
are used, appears significantly above $1.65~$GeV. This observation agrees with the previous lattice results, 
where the quark fields are realized with clover fermions, using only $qqq$ type interpolators. 
It is also to be noted that no interpolator basis renders  more than three eigenstates below 
$1.65~$GeV. This indicates not only that the $qqq$ type interpolators have weak coupling 
with the scattering levels, but also the first excited state with a strong $qqq$ Fock component 
appears significantly above $1.65~$GeV. Considering the closeness of the first and second 
excitations to the non-interacting energies of the scattering levels, we relate them with the 
scattering channels $N(0)\pi(0)\pi(0)$ and the $N(1)\pi(-1)$ respectively. This identification is 
supported by features of the spectrum for different choices of the operator basis as 
demonstrated in Fig. \ref{fig:E_interp_set}. Exclusion of $O^{N(0)\sigma(0)}$ from basis 1 leads to the 
disappearance of first excited state as can be seen in basis 4. Similarly the $N(1)\pi(-1)$ 
Fock component is observed to be crucial to determine the second excitation in basis 1, as can 
be inferred from basis 3.

\begin{figure*}[!htb]
\centering
\sidecaption
\includegraphics*[height=72mm,clip]{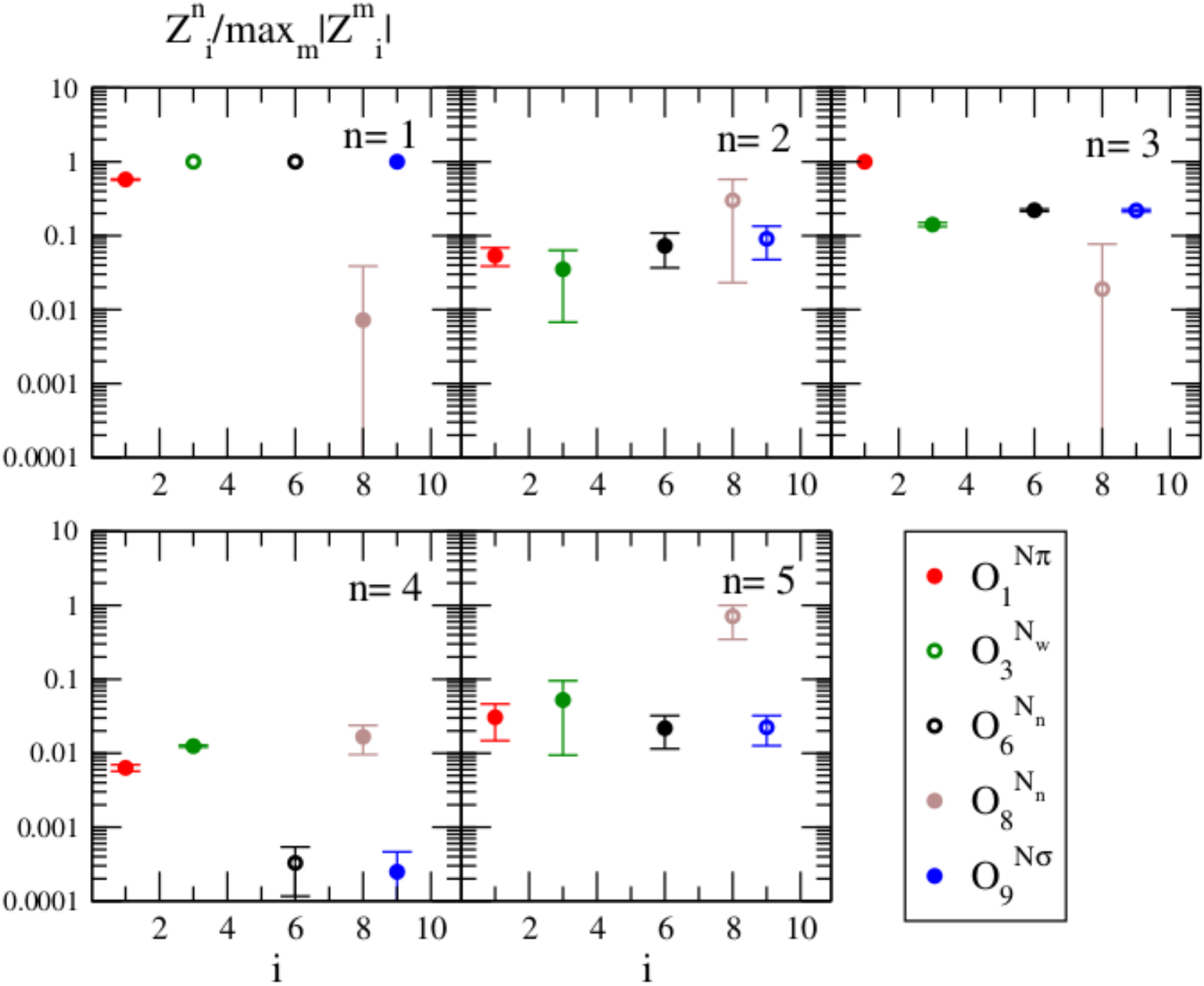}
\caption{The normalized overlaps $\tilde Z_i^n=\langle \Omega| O_i|n\rangle$, extracted from the correlation 
matrix (\ref{C}) based on the complete set (basis 1). The  ratios of overlaps $Z_i^n$ are with respect 
to  the largest among $|Z_i^{m=1,...5}|$ for a given operator $i$. }
\label{fig:E_final}
\end{figure*}

Identification of the dominant nature of the levels from above procedures is further verified by 
their signatures in the overlap factors, $Z_i^n$. We present the normalized overlap factors from the basis 
1 in Fig. \ref{fig:E_final}. The overlap factors $Z_i^n$ for any operator $i$ are normalized with 
the largest among $|Z_i^{m=1,...5}|$. The ground state is observed to couple strongly to all 
interpolators. The operator $O^{N\pi}$ can be seen to overlap strongest with the second excitation, 
implying that the state is related to $N(1)\pi(-1)$. $O^{N\sigma}$ couples strongest to the ground 
state, which could be a result of the $\sigma$ meson mixing with the vacuum. However, the most interesting 
observation in this figure is that the $O^{N\sigma}$ is found to have similar overlap with 
first and second excited states. Furthermore $|\langle \Omega| O^{N\sigma} |n=2\rangle| \lesssim  |\langle \Omega|O^{N\sigma} |n=3\rangle|$  
suggests that the scattering channels could be significantly coupled.

Extracting the phase shift information is not straightforward as the Roper channel is significantly 
inelastic above the $N\pi\pi$ threshold. Numerical study of three-body channels has not been performed 
in lattice QCD up to now, though analytic treatments exist \cite{Hansen:2015zga,Hansen:2016ync}. Hence 
we restrict ourselves to the elastic approximation of $N\pi$ scattering. Within this approximation, 
we determine the infinite volume phase shifts from the lattice energy levels using L\"uscher's relation 
as in eqn. (\ref{luscher}). The energy level related to $N(1)\pi(-1)$ in basis 1 lies on top of the 
non-interacting energy $E_{N(1)}+E_{\pi(1)}$, as can be seen from basis 1 in Figs. \ref{fig:E_interp_set}. 
Hence the energy shift of this level with respect to the non-interacting level and the corresponding infinite 
volume phase shift are consistent with zero within the sizable errors. This observation remains stable 
with respect to various choices of dispersion relation to determine the non-interacting levels, 
correlated or uncorrelated fits, etc. This indicates that $N\pi$ scattering in the elastic approximation 
alone does not render a low lying Roper resonance. 

\begin{figure}[htb]
\centering
\sidecaption
\includegraphics[width=63mm,clip]{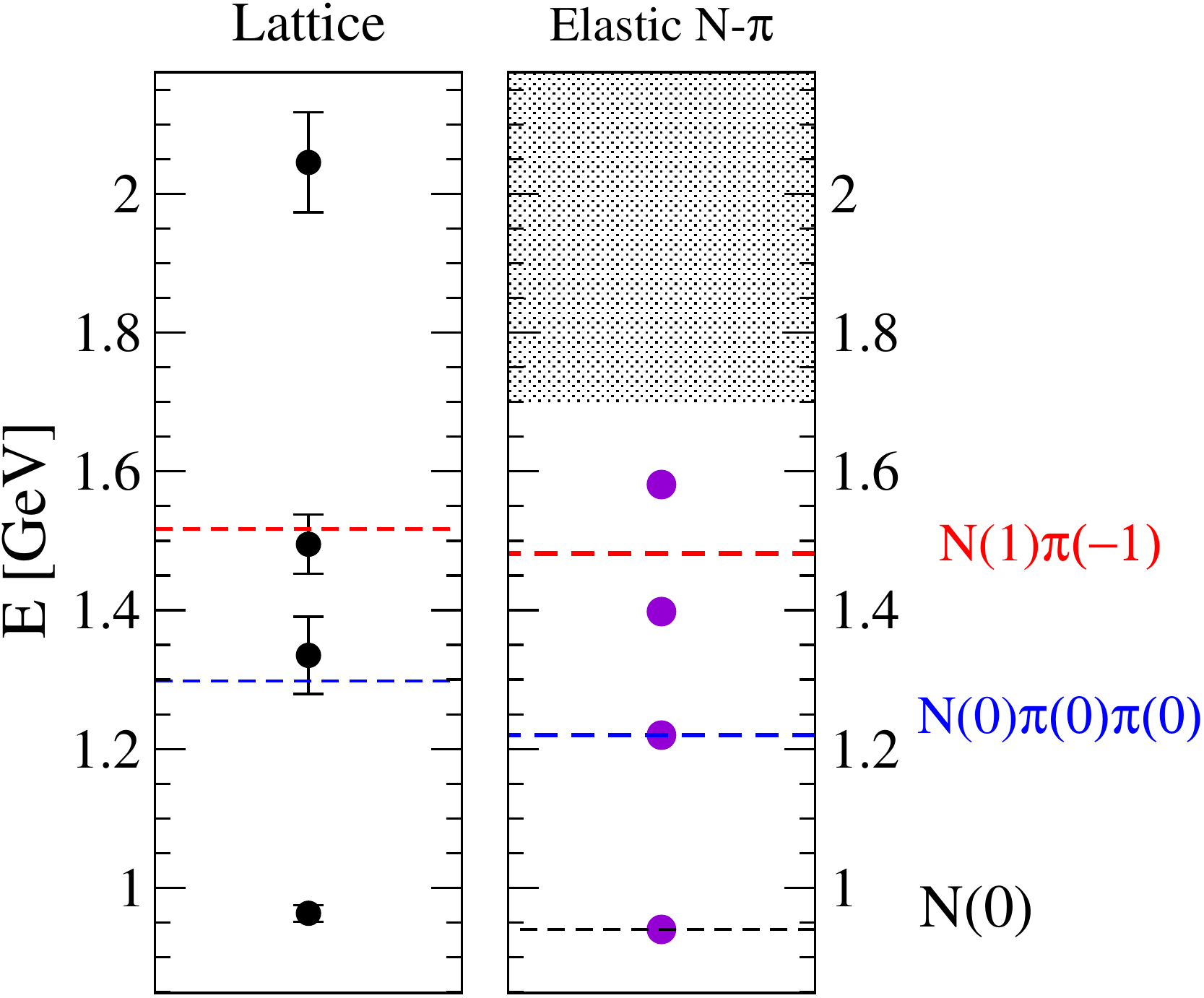}
\caption{Energy values from our simulation (on the left pane) being compared with analytic 
predictions based on L\"uscher's formulae (on the right pane). The relevant non-interacting 
scattering levels are shown as dashed horizontal lines. }
\label{fig:E_analytic}
\end{figure}

We quantify this inference further by comparing the actual lattice data we compute with analytical predictions 
for lattice data in Fig. \ref{fig:E_analytic}. These analytic predictions are based on L\"uscher's formulae 
(assuming elasticity) with experimental inputs on the masses and the phase shifts related to the Roper 
resonance. One can clearly see that below $E<1.65~$GeV numerical results indicate only three eigenlevels, 
whereas analytic predictions indicate additional levels if the Roper can be described as a conventional 
resonance within the elastic $N\pi$ scattering approximation. 

Absence of signatures for the Roper resonance could be due to various reasons. 
The Roper could be a dynamically coupled channel phenomenon involving multiple scattering channels 
like $\Delta \pi$, N$\rho$ and $N\pi\pi$ in the interpolator basis. Or it could be a hybrid 
phenomenon, where the gluonic degrees of freedom are excited \cite{Golowich:1982kx,Kisslinger:1995yw}. 
It could also be that our basis lack some genuine interpolators like a pentaquark operator or 
non-local $qqq$ type interpolators, which could actively scan radial and orbital excitations 
within the $qqq$ structure. Lastly this work, like most other lattice studies, involved quark 
fields that are not chiral at finite lattice spacing $a$. This is probably reflected in the contrasting 
results from Ref. \cite{Liu:2014jua} with respect to other lattice investigations, including ours.

In Ref. \cite{Kiratidis:2016hda} a local pentaquark operator with color structure 
$\epsilon_{abc} \bar q_a [qq]_b[qq]_c$ ($[qq]_c=\epsilon_{cde} q_cq_dq_e$) was used to explore the Roper 
region, but a low-lying Roper state was not found. The local pentaquark 
operator are related to baryon-meson interpolators via Fierz relations, with the important ones 
being the  $N(1)\pi(-1)$ and $N(0)\sigma(0)$ in the Roper region. Since we include such 
baryon-meson interpolators with good partial wave projection, we expect that our simulation 
does incorporate good localized pentaquark operators as well. However, it remains to be a challenge 
to include pentaquark interpolators with more complicated structures. 

Interpolators of type $qqq$ with particularly designed non-local internal structure might also be important 
and be better suited to scan the radial and orbital excitations rather than interpolators with varying 
smearing widths. Most of the studies in the past with such non-local $qqq$ operators have also not yielded 
low-lying state with mass below 1.6 GeV \cite{Edwards:2011jj}. However no work has been reported studying 
the Roper channel with a large number of specially-designed non-local $qqq$ operators as well as relevant 
baryon-meson interpolators in the basis. 

\begin{figure}[!htb]
\centering
\sidecaption
\includegraphics*[width=0.49\textwidth,clip]{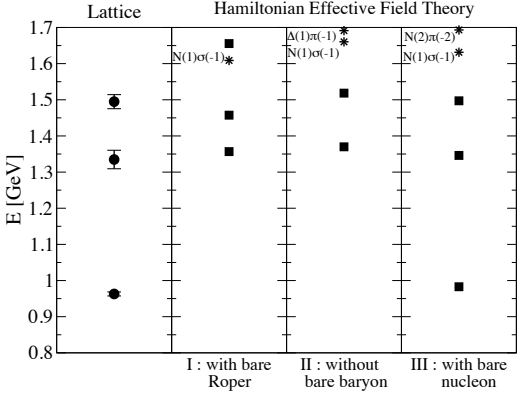}
\caption{Comparison of our results to predictions based on HEFT \cite{Liu:2016uzk} with different Hamiltonians.
Our results are consistent with the scenarios (II) and (III), whereas (I) is disfavored, as it predicts 
additional levels in the energy region being investigated.}
\label{fig:HEFT}
\end{figure}

Dynamically coupled channel phenomena to describe the Roper resonance have been explored recently in 
the \cite{Liu:2016uzk} using Hamiltonian Effective Field Theory (HEFT). Various scenarios involving 
(I) $N\pi-N\sigma-\Delta\pi$ coupled to a low-lying bare $qqq$ type interpolator representing the Roper, 
(II) coupling only to $N\pi-N\sigma-\Delta\pi$ channels and (III) $N\pi-N\sigma-\Delta\pi$ coupled to
a low-lying bare $qqq$ type interpolator representing nucleon were explored using HEFT to explain 
the experimentally observed phase shifts. Figure \ref{fig:HEFT} depicts the comparison of our lattice 
spectrum with the lattice spectrum predicted from the above HEFT calculation. Excluding the stars in 
the spectrum, which are related to the scattering levels that are not included in basis used in our work, 
we compare the number of expected levels (squares and circles) between the lattice and the predictions 
based on HEFT. Our results are compatible with the cases (II) and (III), where one can see that there 
are only two states in the region between 1.2 to 1.7 GeV. However, scenario (I) is disfavored by our 
results as it predicts three levels in this energy range. A recent follow up HEFT calculation, based 
on constraints from our lattice QCD data, has claimed that the Roper resonance is best described as 
a dynamical phenomenon through strongly coupled baryon-meson channels \cite{Wu:2017qve}.

\section{Summary}\label{sec:Conc}
 
In this talk we present the results from our recent lattice QCD study of $N\pi$ scattering in the positive-parity
nucleon channel. This channel is interesting for the presence of a low lying first excitation of the nucleon, 
called Roper resonance  $N^*(1440)$. Employing a set of interpolators that includes $qqq$-like, $N\pi$ in 
$p$-wave and $N\sigma$ in $s$-wave, we systematically extract the excited nucleon spectrum up to an energy of 
1.65 GeV. The extracted lattice spectrum disfavors the description of the  low lying Roper resonance as 
a conventional resonance in the N$\pi$ scattering within the elastic approximation. The overlap factors 
from our study indicate a possibility of Roper being a dynamical coupled channel phenomenon. Other possible 
reasons for absence of signature of Roper resonance in our calculations were also discussed. 

\section{Acknowledgements}
 
We thank the PACS-CS collaboration for providing the gauge configurations.  This work is
supported in part by the  Slovenian Research Agency ARRS, by the Austrian Science Fund
FWF:I1313-N27 and by  the Deutsche Forschungsgemeinschaft Grant No. SFB/TRR 55. The
calculations were performed on  computing clusters at the University of Graz (NAWI Graz)
and Theoretical Department at Jozef Stefan Institute, Ljubljana. M. P. acknowledges 
support from EU under grant no. MSCA-IF-EF-ST-744659 (XQCDBaryons).

\clearpage
\bibliography{lattice2017}

\end{document}